\documentclass{article}

\PassOptionsToPackage{numbers, compress}{natbib}

\usepackage[dblblindworkshop, final]{neurips_2026}

\usepackage[utf8]{inputenc} 
\usepackage[T1]{fontenc}    
\usepackage{hyperref}       
\usepackage{url}            
\usepackage{booktabs}       
\usepackage{amsfonts}       
\usepackage{nicefrac}       
\usepackage{microtype}      
\usepackage{xspace}         
\usepackage{amsmath}        
\usepackage{xcolor}         
\usepackage{graphicx}       

\usepackage{tikz}
\usetikzlibrary{positioning, arrows.meta, calc, fit, decorations.pathreplacing, backgrounds}

\usepackage{bm}
\definecolor{bestblue}{HTML}{0B5394}
\newcommand{\best}[2]{\textcolor{bestblue}{\textbf{#1\textsubscript{\textpm#2}}}}

\newcommand{\mboformer}{\textsf{MBOFormer}\xspace}  
\newcommand{\mbofusion}{\textsf{MBOFusion}\xspace}

\title{Event History Over Scale: Compact Transformers for Low-Latency Limit Order Book Forecasting}

\workshoptitle{TS-LIMITS}

\author{%
  \textbf{David Schaurecker}$^{1,}$\thanks{Equal contribution.\\The code is available at \url{https://github.com/lassebaerlandstrand/event-history-over-scale}.}, \quad
  \textbf{Lasse B. Strand}$^{1,}$\footnotemark[2], \quad
  \textbf{Kevin O'Sullivan}$^{1}$, \quad
  \textbf{Robert Jakob}$^{1}$ \\[0.8ex]
  $^{1}$Dep. of Management, Technology, and Economics, ETH Zurich, Switzerland \\[0.8ex]
  \texttt{schaureckerd@ethz.ch}, \quad
  \texttt{lstrand.research@gmail.com}
}

\begin{document}

\maketitle

\begin{abstract}
Short-horizon price-trend prediction from limit order books in equity and intraday electricity markets requires models that combine predictive quality with low single-sample latency and a small serialized model size to keep pace with rapid and continuous market updates. We introduce MBOFormer, a 7{,}203-parameter causal transformer, and MBOFusion, a 14{,}371-parameter extension with a slow temporal-context branch. Both models process market-by-order (level-3) histories of individual order submissions, cancellations, and executions. We compare them with level-2-based baselines that process sampled order-book snapshots and simple statistics instead of their underlying messages. Across three markets and four prediction horizons, our models achieve the highest mean macro-F1 in eleven of the twelve settings in a comparison against other benchmark models of varying sizes. Both our models achieve sub-millisecond median inference on a single Apple M2 CPU thread and have serialized state dictionaries below 70\,kB. MBOFormer is $5.1\times$--$8.1\times$ faster than the million-parameter baselines at comparable predictive quality. Our results show that fine-grained market event history can reduce the need for model size under tight deployment constraints for short-term equity and electricity price forecasting tasks.
\end{abstract}

\section{Introduction}
\label{sec:intro}
Short-horizon price-trend prediction from limit order book (LOB) data is a core problem in market microstructure \cite{ntakaris2018fi2010, kercheval2015modelling}. Deploying these models at the market's rapid update rate also makes inference latency and memory major practical challenges. In this setting, the time series is not an evenly spaced sequence of prices but an irregular sequence of limit order submissions, cancellations, and executions whose cumulative effect updates the order book and produces the observed mid-price. Because predictions are made online at individual decision points where batching would add a delay, we evaluate batch-one latency alongside memory footprint. Latency generally depends on parameter count, sequence length, operator selection, hardware, and runtime overhead \cite{menghani2023efficient}. Moreover, because deployed models are queried continuously at every decision point, small per-inference costs accumulate, motivating reporting computational demands alongside predictive quality \cite{schwartz2020green, henderson2020systematic}.

\definecolor{lobNavy}{HTML}{17324D}
\definecolor{lobSlate}{HTML}{F1F4F7}
\definecolor{lobBlue}{HTML}{DCEBFA}
\definecolor{lobBlueDark}{HTML}{3E78A5}
\definecolor{lobRed}{HTML}{F9E2E0}
\definecolor{lobRedDark}{HTML}{B85C59}
\definecolor{lobTeal}{HTML}{DDF3ED}
\definecolor{lobTealDark}{HTML}{3F806F}
\definecolor{lobGold}{HTML}{FFF0CF}
\definecolor{lobGoldDark}{HTML}{9A7318}
\definecolor{lobPanel}{HTML}{FAFBFC}

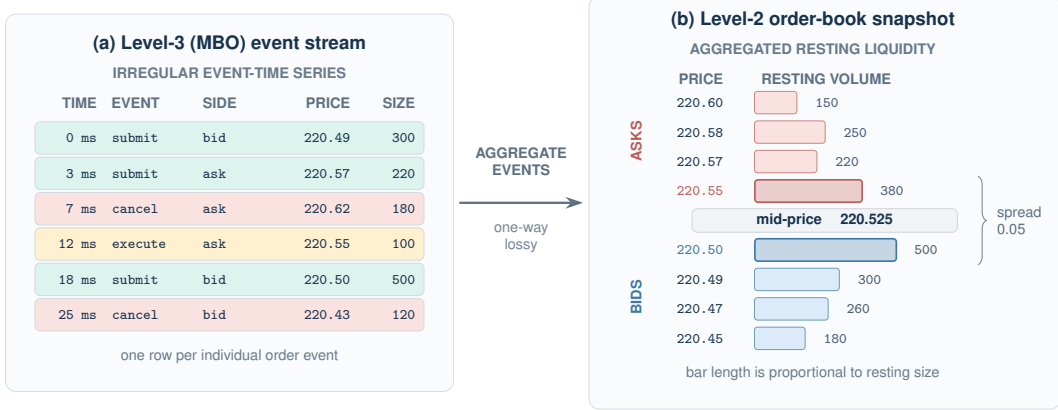
\begin{figure}[t]
\centering
\resizebox{\textwidth}{!}{%
\begin{tikzpicture}[
  x=1mm,
  y=1mm,
  panel/.style={draw=lobNavy!16, fill=lobPanel, rounded corners=5pt, line width=0.5pt},
  title/.style={font=\small\sffamily\bfseries, text=lobNavy},
  subtitle/.style={font=\scriptsize\sffamily\bfseries, text=lobNavy!65},
  header/.style={font=\scriptsize\sffamily\bfseries, text=lobNavy!70},
  eventrow/.style={draw=lobNavy!18, line width=0.4pt, rounded corners=2.5pt, minimum width=66mm, minimum height=5.5mm, font=\scriptsize\ttfamily, text=lobNavy},
  submit/.style={eventrow, fill=lobTeal},
  cancel/.style={eventrow, fill=lobRed},
  execute/.style={eventrow, fill=lobGold},
  price/.style={anchor=east, font=\scriptsize\ttfamily, text=lobNavy},
  bestprice/.style={price, font=\scriptsize\ttfamily\bfseries},
  size/.style={anchor=west, font=\scriptsize\ttfamily, text=lobNavy!80},
  askbar/.style={draw=lobRedDark!75, fill=lobRed, rounded corners=1.5pt, minimum height=3.8mm, inner sep=0pt},
  bidbar/.style={draw=lobBlueDark!75, fill=lobBlue, rounded corners=1.5pt, minimum height=3.8mm, inner sep=0pt},
  bestask/.style={askbar, draw=lobRedDark, fill=lobRedDark!30, line width=0.8pt},
  bestbid/.style={bidbar, draw=lobBlueDark, fill=lobBlueDark!30, line width=0.8pt},
  midband/.style={draw=lobNavy!20, fill=lobSlate, rounded corners=2.5pt, minimum height=5.5mm, align=center, font=\scriptsize\sffamily\bfseries, text=lobNavy},
  flow/.style={-{Stealth[length=3mm,width=2mm]}, draw=lobNavy!70, line width=1.1pt},
]
  \begin{scope}[on background layer]
    \node[panel, minimum width=76mm, minimum height=64mm] at (38,-25) {};
    \node[panel, minimum width=80mm, minimum height=70mm] at (139,-25) {};
  \end{scope}

  \begin{scope}[yshift=-4mm]
    \node[title] at (38,6) {(a) Level-3 (MBO) event stream};
    \node[subtitle] at (38,1) {IRREGULAR EVENT-TIME SERIES};
    \node[header] at (38,-4) {\makebox[9mm][r]{TIME}\makebox[18mm][l]{\quad EVENT}\makebox[10mm][l]{SIDE}\makebox[15mm][r]{PRICE}\makebox[11mm][r]{SIZE}};
    \node[submit] at (38,-10) {\makebox[9mm][r]{0 ms}\makebox[18mm][l]{\quad submit}\makebox[10mm][l]{bid}\makebox[15mm][r]{220.49}\makebox[11mm][r]{300}};
    \node[submit] at (38,-16) {\makebox[9mm][r]{3 ms}\makebox[18mm][l]{\quad submit}\makebox[10mm][l]{ask}\makebox[15mm][r]{220.57}\makebox[11mm][r]{220}};
    \node[cancel] at (38,-22) {\makebox[9mm][r]{7 ms}\makebox[18mm][l]{\quad cancel}\makebox[10mm][l]{ask}\makebox[15mm][r]{220.62}\makebox[11mm][r]{180}};
    \node[execute] at (38,-28) {\makebox[9mm][r]{12 ms}\makebox[18mm][l]{\quad execute}\makebox[10mm][l]{ask}\makebox[15mm][r]{220.55}\makebox[11mm][r]{100}};
    \node[submit] at (38,-34) {\makebox[9mm][r]{18 ms}\makebox[18mm][l]{\quad submit}\makebox[10mm][l]{bid}\makebox[15mm][r]{220.50}\makebox[11mm][r]{500}};
    \node[cancel] at (38,-40) {\makebox[9mm][r]{25 ms}\makebox[18mm][l]{\quad cancel}\makebox[10mm][l]{bid}\makebox[15mm][r]{220.43}\makebox[11mm][r]{120}};
    \node[font=\scriptsize\sffamily, text=lobNavy!65] at (38,-47) {one row per individual order event};
  \end{scope}

  \node[header, align=center] at (87.5,-18) {AGGREGATE\\EVENTS};
  \draw[flow] (77,-25) -- (98,-25);
  \node[font=\scriptsize\sffamily, text=lobNavy!60, align=center] at (87.5,-31) {one-way\\lossy};

  \node[title] at (137,6) {(b) Level-2 order-book snapshot};
  \node[subtitle] at (137,1) {AGGREGATED RESTING LIQUIDITY};
  \node[header, anchor=east] at (123,-4) {PRICE};
  \node[header, anchor=west] at (127,-4) {RESTING VOLUME};

  \node[header, text=lobRedDark, rotate=90] at (107,-14.5) {ASKS};
  \foreach \y/\px/\len/\sz in {-8/220.60/7.2/150, -13/220.58/12.0/250, -18/220.57/10.6/220}{
    \node[price] at (123,\y) {\px};
    \node[askbar, anchor=west, minimum width=\len mm] at (127,\y) {};
    \node[size] at (129+\len,\y) {\sz};
  }
  \node[bestprice, text=lobRedDark] at (123,-23) {220.55};
  \node[bestask, anchor=west, minimum width=18.2mm] at (127,-23) {};
  \node[size] at (147.2,-23) {380};

  \node[midband, minimum width=45mm, minimum height=4mm, inner ysep=0.5mm] at (139,-28) {mid-price \quad 220.525};

  \node[bestprice, text=lobBlueDark] at (123,-33) {220.50};
  \node[bestbid, anchor=west, minimum width=24mm] at (127,-33) {};
  \node[size] at (153,-33) {500};
  \foreach \y/\px/\len/\sz in {-38/220.49/14.4/300, -43/220.47/12.5/260, -48/220.45/8.6/180}{
    \node[price] at (123,\y) {\px};
    \node[bidbar, anchor=west, minimum width=\len mm] at (127,\y) {};
    \node[size] at (129+\len,\y) {\sz};
  }
  \node[header, text=lobBlueDark, rotate=90] at (107,-41) {BIDS};

  \draw[decorate, decoration={brace, amplitude=4pt}, draw=lobNavy!60, line width=0.5pt] (165,-20.5) -- (165,-35.5)
    node[midway, anchor=west, xshift=2mm, font=\scriptsize\sffamily, text=lobNavy!70, align=left]{spread\\0.05};
  \node[font=\scriptsize\sffamily, text=lobNavy!65] at (137,-54) {bar length is proportional to resting size};
\end{tikzpicture}%
}
\caption{The same market evolution represented at two levels of detail. The Level-3 market-by-order stream records each submission, cancellation, and execution in event time. Aggregating these messages produces the Level-2 snapshot of resting volume by price level. This mapping is one-way: the snapshot does not retain the identity, timing, type, or ordering of the events that produced it.}
\label{fig:l2l3}
\end{figure}

Most recent deep LOB models operate on a sampled level-2 (L2) limit order book state, optionally augmented with the triggering message and engineered features. Figure~\ref{fig:l2l3} explains this and contrasts that representation with the market-by-order (MBO, or level-3, L3) stream from which it is derived. An L2 snapshot records aggregated resting volumes at selected price levels of the current limit order book, whereas the L3 stream records each submission, cancellation, and execution message that is processed by the exchange, and is thus the most granular representation available. Replaying L3 data naturally reconstructs L2 data, but a sequence of snapshots cannot recover the identity, timing, order, or type of the intervening messages between LOB snapshots and decision points. Market-microstructure theory motivates a connection between order flow and price formation \cite{kyle1985continuous, ohara2015high}. Empirical work relates order-flow imbalance to concurrent and cross-asset price changes \cite{cont2014priceimpact, cont2023crossimpact}. In this work, we explore whether a compact mid-price forecasting model can leverage this granular information more effectively than rich state and feature histories.

While short-horizon price prediction using L2 snapshots is well studied, comparatively few works exploit granular L3 event data. Among L2 approaches, DeepLOB combines convolutions over the price ladder with a recurrent head \cite{zhang2019deeplob}. Subsequent approaches include bilinear and attention models \cite{tran2019tabl, tran2021bin}, transformer and axial-attention architectures \cite{wallbridge2020translob, kisiel2022axiallob}, and the dual-axis attention model TLOB \cite{berti2025tlob}. For a recent compute- and latency-focused comparison of LOB models on the FI-2010 dataset \cite{ntakaris2018fi2010}, see \cite{hedges2026inference}. Generally, predictability is present but strongly representation-dependent, generally weakens at longer horizons, and need not translate into actionable trading signals \cite{lucchese2024shortterm, briola2025deep}. On the event side, order-flow models identify cross-stock regularities \cite{sirignano2019universal} and gains over book states \cite{kolm2023deepofi}, while prior MBO work finds that message and LOB inputs provide complementary information \cite{zhang2021mbo}. Recent work models granular event streams at substantially larger scales. MarketGPT pretrains an approximately 100-million-parameter autoregressive model on 91 million messages, while TradeFM trains a 524-million-parameter foundation model on billions of trade events \cite{wheeler2024marketgpt, kawawa2026tradefm}. Other generative approaches apply diffusion models either to order flows or to sequences of L2 book states \cite{berti2025trades, backhouse2025painting}. Of particular relevance, LOBERT adapts an encoder-only transformer to sequences of 512 L3 messages and can additionally incorporate contemporaneous L2 snapshots \cite{linna2025lobert}. It is pretrained on 470 million messages and contains approximately 1.1 million parameters, making it 77--153 times larger than our models. LOB-based forecasting is well established for equities but remains comparatively limited for European continuous intraday electricity markets. Existing work studies market dynamics through simulation and point-process models \cite{martin2018german, vonluckner2021hawkes}, forecasts price indices from sequences of matched trades \cite{yu2025orderfusion}, forecasts prices from engineered order-book features \cite{yu2026orderbook}, or predicts price direction from LOB features \cite{hornek2025directional}. We instead ask whether a compact, directly supervised model that retains the most recent raw order events can match or outperform larger models built on coarser, sampled order-book histories while reducing inference latency relative to those larger baselines.

We answer this question with MBOFormer, a 7{,}203-parameter causal transformer over the 128 most recent order events, and MBOFusion, which adds a convolutional branch for slow market context at 14{,}371 parameters. We compare them with established L2-based LOB baselines, including a size-matched and a million-parameter TLOB, on NASDAQ and NYSE equities and the EPEX continuous intraday electricity market. To gauge applicability to a live high-frequency environment, we measure batch-one model-forward latency on precomputed inputs, serialized state-dictionary size, peak activation memory, and estimated FLOPs per forward pass, the last as a hardware-independent proxy for accumulated compute. We do not measure energy or carbon impact directly.

We make three main contributions:
\begin{enumerate}
\setlength{\itemsep}{0pt}
\setlength{\parsep}{0pt}
\setlength{\topsep}{2pt}
    \item We compare our L3-based models with L2-based baselines across three continuous markets, four forecast horizons, and matched model parameter budgets. One of our models has the highest mean macro-F1 among evaluated models in eleven of twelve settings, and MBOFormer achieves a higher mean score than the equally sized TLOB in the same eleven settings.
    \item We measure rather than assume deployment efficiency on one Apple M2 CPU thread. Both of our L3-based models achieve sub-millisecond median inference latency. MBOFormer achieves a $5.1\times$--$8.1\times$ speedup over the million-parameter baselines at comparable prediction quality, while BiNCTABL and the parameter-matched TLOB are faster still but generally sacrifice prediction quality.
    \item We provide the first deep-learning treatment of EPEX electricity exchange order-message histories for short-horizon price forecasting, demonstrating that a compact architecture generalizes across electricity and equity markets.
\end{enumerate}

\section{Method}
\label{sec:method}

\subsection{Prediction task, labels, and event representation}\label{sec:pred_task}
The decision points lie on an event-time grid obtained by sampling the order book after every tenth book-changing message. At each decision point $t$, we observe the market history up to and including $t$ and predict the direction of the mid-price $h \in \{10, 20, 50, 100\}$ sampled steps ahead. Following the event-based convention used in the FI-2010 benchmark \cite{ntakaris2018fi2010}, these horizons are indexed in sampled steps rather than wall-clock time. The L2-based baselines receive recent sampled sequences of L2 states and, where configured, triggering messages and engineered market features, while our L3-based models receive the recent raw L3 message history. MBOFusion additionally receives a slow context vector. Both representations use the same decision points and targets. Prior LOB forecasting work reduces label noise by smoothing mid-prices, although the precise conventions differ \cite{ntakaris2018fi2010, zhang2019deeplob, lucchese2024shortterm}. We use the fixed two-window construction implemented for TLOB \cite{berti2025tlob}: a trailing ten-step reference mean and a ten-step mean ending $h$ sampled steps later. We hold the smoothing length fixed across forecast horizons so that only the forecast distance changes. Let $m_t$ denote the mid-price at sampled step $t$. We define the trailing ten-step mean and signed change at horizon $h$ as
\begin{equation}
\bar{m}_t=\frac{1}{10}\sum_{j=0}^{9}m_{t-j},
\qquad
\Delta_{t,h}=\bar{m}_{t+h}-\bar{m}_t.
\end{equation}
For each horizon $h$, we compute the adaptive threshold over all valid positions in the sequence being labeled:
\begin{equation}
\alpha_h=\frac{1}{2|\mathcal{V}_h|}\sum_{t\in\mathcal{V}_h}|\Delta_{t,h}|,
\end{equation}
where for a sequence of $N$ sampled steps, $\mathcal{V}_h:=\{t:9\leq t\leq N-h-1\}$ contains the positions at which both the current window $(m_{t-9},\ldots,m_t)$ and the future window $(m_{t+h-9},\ldots,m_{t+h})$ lie entirely within the sequence. This adapts TLOB's preprocessing \cite{berti2025tlob} to signed price differences, accommodating zero or negative electricity prices. For equities, the labeled sequence is each complete training, validation, or test split, whereas for EPEX, each delivery contract is assigned to one of these splits and labeled as a separate sequence. We label an observation \textit{up} when $\Delta_{t,h}>\alpha_h$, \textit{down} when $\Delta_{t,h}<-\alpha_h$, and \textit{flat} otherwise. Because $\alpha_h$ is computed from all valid positions in each sequence, the test labels use the full test sequence's price-change distribution. This is a look-ahead in the label definition, not in any model input. The resulting labels are applied identically to all compared models, but a live system would have to fix $\alpha_h$ from past data, so absolute scores should not be read as achievable online.

Each of our L3-based models receives a right-aligned window of up to $T=128$ recent events. Replaying the message stream yields seven per-event features: (1) event type (submission, cancellation, execution), (2) signed side ($\pm 1$), (3) log tick-depth from the current best quote, (4) log message size, (5) log elapsed time, (6) immediate mid-price change, and (7) a best-quote indicator. All features are solely derived from the L3 stream without separate L2 inputs.

\subsection{Our models}
\textbf{\mboformer} (Figure~\ref{fig:arch}a) is a compact causal transformer that processes the $T$ most recent L3 events defined above at each decision point. A tokenizer maps each event to a $d$-dimensional token by embedding the event type, linearly projecting the remaining fields, and zeroing out padded positions. We add a learned positional embedding and apply two transformer blocks, each comprising a multi-head self-attention layer under a causal mask followed by a position-wise feed-forward network \cite{vaswani2017attention}. The causal mask lets every token attend only to earlier events, so the representation at the final position summarizes the market as of the decision time with no look-ahead. After a final layer normalization, we pool the sequence by concatenating the newest-event representation with the mask-aware mean over the valid events, and a two-layer perceptron maps the pooled vector to our three prediction classes. At a hidden width of $d = 16$, with two blocks and four attention heads, MBOFormer has 7{,}203 parameters. The small width reflects the input: each event is summarized by only a few informative fields, so the encoder needs little capacity once the representation exposes the order-flow signal. Because every decision consumes a fixed window of the $T$ most recent events and the model carries no state between decisions, its memory footprint is fixed by $T$ and does not grow with the length of the order stream.

\textbf{\mbofusion} (Figure~\ref{fig:arch}b) extends MBOFormer with a slow temporal-context branch and reuses the event encoder unchanged. Alongside the event window we maintain a 12-dimensional context vector, sampled once per step on the grid of Section~\ref{sec:pred_task}, which summarizes recent market conditions. It includes mid-price drift and volatility, average spread, the depth of resting liquidity, event and execution rates, the balance of submissions to cancellations, and, where the market defines one, the time remaining until the contract stops trading. The context branch projects these vectors, applies one causal convolution over the 16 most recent sampled steps, and concatenates the final, mean, and maximum activations. We concatenate this context summary with the pooled event representation and classify from the joint vector. MBOFusion thus has 14{,}371 parameters. Because the branch sits on top of MBOFormer's unchanged event encoder, their comparison evaluates the combined effect of additional temporal context and the parameters introduced by the context branch.
\definecolor{archNavy}{HTML}{17324D}
\definecolor{archSlate}{HTML}{F1F4F7}
\definecolor{archBlue}{HTML}{DCEBFA}
\definecolor{archBlueStrong}{HTML}{DCEBFA}
\definecolor{archTeal}{HTML}{DDF3ED}
\definecolor{archGold}{HTML}{FFF0CF}
\definecolor{archViolet}{HTML}{EEE7F7}
\definecolor{archPanel}{HTML}{FAFBFC}

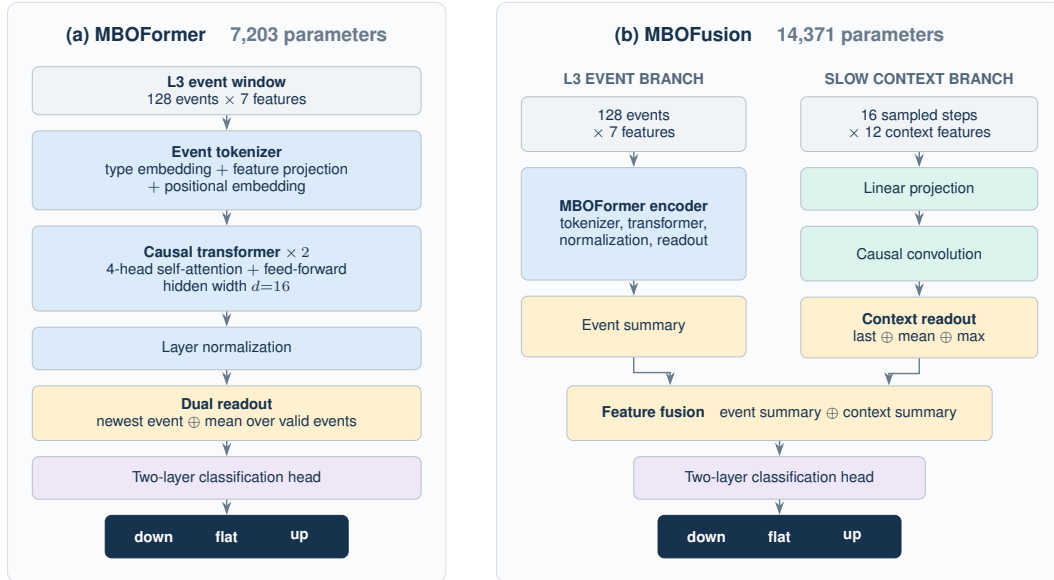
\begin{figure}[h!]
\centering
\resizebox{\textwidth}{!}{%
\begin{tikzpicture}[
  x=1mm,
  y=1mm,
  card/.style={draw=archNavy!30, line width=0.45pt, rounded corners=3pt, align=center, font=\scriptsize\sffamily, text=archNavy, minimum height=7mm, inner xsep=3mm, inner ysep=1.5mm},
  input/.style={card, fill=archSlate},
  event/.style={card, fill=archBlue},
  eventstrong/.style={card, fill=archBlueStrong},
  context/.style={card, fill=archTeal},
  aggregate/.style={card, fill=archGold},
  classifier/.style={card, fill=archViolet},
  output/.style={card, fill=archNavy, draw=archNavy, text=white, font=\scriptsize\sffamily\bfseries},
  title/.style={font=\small\sffamily\bfseries, text=archNavy, inner sep=0pt},
  branchlabel/.style={font=\scriptsize\sffamily\bfseries, text=archNavy!75},
  flow/.style={-{Stealth[length=2.2mm,width=1.5mm]}, draw=archNavy!65, line width=0.65pt},
  panel/.style={draw=archNavy!16, fill=archPanel, rounded corners=5pt, line width=0.5pt},
]
  \node[title] (atitle) at (39,6) {(a) MBOFormer \quad \textcolor{archNavy!65}{7{,}203 parameters}};
  \node[input, minimum width=64mm] (ain) at (39,-3) {\textbf{L3 event window}\\128 events $\times$ 7 features};
  \node[event, minimum width=64mm, minimum height=13mm, below=2.5mm of ain] (atok) {\textbf{Event tokenizer}\\type embedding $+$ feature projection\\$+$ positional embedding};
  \node[eventstrong, minimum width=64mm, minimum height=14mm, below=2.5mm of atok] (atx) {\textbf{Causal transformer $\times\,2$}\\4-head self-attention $+$ feed-forward\\hidden width $d{=}16$};
  \node[event, minimum width=64mm, below=2.5mm of atx] (anrm) {Layer normalization};
  \node[aggregate, minimum width=64mm, minimum height=9mm, below=2.5mm of anrm] (apool) {\textbf{Dual readout}\\newest event $\oplus$ mean over valid events};
  \node[classifier, minimum width=64mm, below=2.5mm of apool] (ahead) {Two-layer classification head};
  \node[output, minimum width=40mm, below=2.5mm of ahead] (aout) {};
  \node[font=\scriptsize\sffamily\bfseries, text=white] at ([xshift=-12mm]aout.center) {down};
  \node[font=\scriptsize\sffamily\bfseries, text=white] at (aout.center) {flat};
  \node[font=\scriptsize\sffamily\bfseries, text=white] at ([xshift=12mm]aout.center) {up};
  \foreach \a/\b in {ain/atok, atok/atx, atx/anrm, anrm/apool, apool/ahead, ahead/aout}
    \draw[flow] (\a) -- (\b);

  \node[title] (btitle) at (130,6) {(b) MBOFusion \quad \textcolor{archNavy!65}{14{,}371 parameters}};
  \node[branchlabel] at (106,-1) {L3 EVENT BRANCH};
  \node[branchlabel] at (153,-1) {SLOW CONTEXT BRANCH};

  \node[input, minimum width=37mm, minimum height=9mm] (bein) at (106,-8.5) {128 events\\$\times$ 7 features};
  \node[input, minimum width=39mm, minimum height=9mm] (bcin) at (153,-8.5) {16 sampled steps\\$\times$ 12 context features};

  \node[context, minimum width=39mm, below=2.5mm of bcin] (bproj) {Linear projection};
  \node[context, minimum width=39mm, minimum height=9mm, below=2.5mm of bproj] (bconv) {Causal convolution};
  \node[aggregate, minimum width=39mm, minimum height=10mm, below=2.5mm of bconv] (bread) {\textbf{Context readout}\\last $\oplus$ mean $\oplus$ max};

  \node[eventstrong, minimum width=37mm, minimum height=18.5mm, below=2.5mm of bein] (benc) {\textbf{MBOFormer encoder}\\tokenizer, transformer,\\normalization, readout};
  \node[aggregate, minimum width=37mm, minimum height=10mm, below=2.5mm of benc] (besum) {Event summary};

  \draw[flow] (bein) -- (benc);
  \draw[flow] (benc) -- (besum);
  \draw[flow] (bcin) -- (bproj);
  \draw[flow] (bproj) -- (bconv);
  \draw[flow] (bconv) -- (bread);

  \node[aggregate, minimum width=70mm, minimum height=9mm] (bcat) at (130, 0 |- apool) {\textbf{Feature fusion}\quad event summary $\oplus$ context summary};
  \node[classifier, minimum width=48mm, below=2.5mm of bcat] (bhead) {Two-layer classification head};
  \node[output, minimum width=40mm, below=2.5mm of bhead] (bout) {};
  \node[font=\scriptsize\sffamily\bfseries, text=white] at ([xshift=-12mm]bout.center) {down};
  \node[font=\scriptsize\sffamily\bfseries, text=white] at (bout.center) {flat};
  \node[font=\scriptsize\sffamily\bfseries, text=white] at ([xshift=12mm]bout.center) {up};

  \draw[flow] (besum.south) -- ++(0,-2.5mm) -| ([xshift=-18mm]bcat.north);
  \draw[flow] (bread.south) -- ++(0,-2.5mm) -| ([xshift=18mm]bcat.north);

  \draw[flow] (bcat) -- (bhead);
  \draw[flow] (bhead) -- (bout);

  \begin{scope}[on background layer]
    \node[panel, fit=(atitle)(ain)(atok)(atx)(anrm)(apool)(ahead)(aout), inner sep=4mm] {};
    \node[panel, fit=(btitle)(bein)(benc)(besum)(bcin)(bproj)(bconv)(bread)(bcat)(bhead)(bout), inner sep=4mm] {};
  \end{scope}
\end{tikzpicture}%
}
\caption{Architectures of our two L3-based models. Blue denotes event-sequence processing, green denotes the slow context branch, gold denotes aggregation or fusion, and purple denotes classification. (a) MBOFormer encodes the 128 most recent order events. (b) MBOFusion reuses the MBOFormer event encoder and adds a branch over the 16 most recent context vectors.}
\label{fig:arch}
\end{figure}

\section{Experimental Setup}
\label{sec:setup}

\subsection{Markets and data}
We evaluate our models across three continuous markets using native L3 market-by-order message data. Among these is the European continuous intraday electricity market, which is an increasingly relevant testbed for latency-constrained execution outside traditional finance applications, where frequent LOB updates and timely reactions to them materially dictate trading policy performance even for asset-backed policies \cite{boukas2021drl, schaurecker2025battery}. Our EPEX SPOT data, obtained directly from the electricity exchange and not publicly available, covers 13 delivery days in March 2021 for hourly delivery products only. We also use L3 NASDAQ AAPL (Apple stock) data over 7 days from 3 to 11 September 2024, and NYSE BAC (Bank of America Corporation stock) data over 58 days from 1 August to 22 October 2024, both obtained through Databento. The EPEX data is comparatively thin, with a median price spread of about 60 price ticks and an average of 13 seconds (median 3 seconds) per sampled event-step for each product. The AAPL and BAC data have median price spreads of around 2 and 1 price ticks, and 0.045 and 0.5 seconds per event-step on average, respectively.

As described in Section~\ref{sec:pred_task}, we sample the order book after every tenth book-changing message rather than on a wall-clock grid. Accordingly, the forecasting horizon $h=10$ spans on average roughly half a second on AAPL, five seconds on BAC, and two minutes on EPEX, where liquidity increases rapidly as delivery approaches \cite{kuppelwieser2021, kremer2021econometric}. We split complete days chronologically into train/validation/test sets: 8/2/3 days for EPEX, 5/1/1 for AAPL, and 35/9/14 for BAC. Every test day therefore postdates every training day, so the reported scores measure out-of-time generalization to later market conditions rather than interpolation within a period. The train/validation/test splits contain approximately 1.00/0.28/0.42 million sampled decision points for EPEX, 2.57/0.42/0.50 million for AAPL, and 1.51/0.37/0.61 million for BAC. To prevent information used to form labels from spanning dataset boundaries, following leakage-aware splitting practices for high-frequency financial data \cite{karpman2023learning}, we construct labels so that no label window crosses a split boundary. Each market provides roughly 1.7--3.5 million sampled steps.

The three markets differ sharply in how often the mid-price moves enough to leave the \textit{flat} class. Under the labeling rule of Section~\ref{sec:pred_task}, the class covers about 70\% of the training windows on the thin electricity market at $h=10$, against 44\% on BAC and 38\% on AAPL, with similar shares on the held-out splits. As the forecast horizon grows, the mid-price crosses the threshold more often, so the \textit{flat} share falls and the three classes move toward balance, reaching roughly 51\% on the electricity dataset and 29--33\% for the equities at $h=100$. This reduces the advantage of always predicting \textit{flat} and makes the prediction task harder at longer horizons.

\subsection{Models and inputs}
We compare our L3-based models, MBOFormer (7{,}203 parameters) and MBOFusion (14{,}371 parameters), with the L2-based baselines DeepLOB \cite{zhang2019deeplob}, BiNCTABL \cite{tran2019tabl, tran2021bin}, MLPLOB, and TLOB \cite{berti2025tlob}. Each sampled L2 snapshot contains 40 values: the price and aggregated volume at ten bid and ten ask levels. DeepLOB processes windows of 100 such snapshots, whereas BiNCTABL processes windows of 10 and MLPLOB windows of 384. DeepLOB and BiNCTABL use only these 40 values. MLPLOB and TLOB receive 14 additional fields per snapshot: six fields for the triggering message and eight engineered order-flow imbalance (OFI) and volume-weighted average price (VWAP) features. We call these models L2-based because their temporal input is a sequence of sampled L2 rows. Although some receive features derived from the triggering L3 message, none receive a history of individual L3 events.

We run TLOB at 1.20M, 23.9k, and 7.2k parameters. The 7{,}206-parameter version matches MBOFormer's size but receives a 16-step sampled window rather than 128 raw events. The 3.14M-parameter MLPLOB (3.07M on the equities) and 1.20M-parameter TLOB test whether greater capacity or a longer sampled history provides a predictive advantage. We exclude LOBERT \cite{linna2025lobert} because its implementation and weights are unavailable and Appendix~\ref{app:baseline-details} documents our TLOB adaptations for this work's pipeline.

\subsection{Training and evaluation}
We train each model with the AdamW optimizer and keep the checkpoint with the lowest validation loss for testing, stopping early when the validation loss fails to decrease by at least $5\times10^{-4}$ for two consecutive epochs. The learning rate and batch size are set per model type and size so that each baseline trains at a stable operating point. Each (model, market, horizon) combination is a separate single-horizon model, so for each tested model, we train $4\times3 = 12$ combinations. We evaluate each configuration across three random initialization seeds and report the test-set macro-F1 score. Macro-averaging calculates the unweighted mean across the three classes (\textit{up}, \textit{flat}, and \textit{down}), preventing the dominant \textit{flat} class from obscuring performance on the rarer \textit{up} and \textit{down} movements, a safeguard given the class imbalance, particularly in the electricity market. A predictor that always outputs \textit{flat} therefore reaches a test macro-F1 of only about 0.26 on EPEX, 0.19 on AAPL, and 0.22 on BAC at $h=10$. Each event-model run trains in at most about twenty minutes per seed on a single NVIDIA GeForce RTX 4080.

\section{Results}
\label{sec:results}

\begingroup
\setlength{\parskip}{3pt}
\makeatletter
\renewcommand{\paragraph}{%
  \@startsection{paragraph}{4}{\z@}%
                {0.6ex \@plus 0.2ex \@minus 0.1ex}%
                {-1em}%
                {\normalsize\bf}%
}
\makeatother

\begin{table}[t]
\centering
\small
\caption{Highest mean test macro-F1 among the L2-based baselines (DeepLOB, BiNCTABL, MLPLOB, and TLOB) and among our L3-based models (MBOFormer and MBOFusion), with the maximizing model selected separately in each cell. Horizons $h$ are sampled steps of ten events each. Full per-model results with standard deviations are given in Appendix~\ref{app:full-results}.}
\label{tab:predictive-summary}
\begin{tabular}{llcccc}
\toprule
Market & Model family & $h{=}10$ & $h{=}20$ & $h{=}50$ & $h{=}100$ \\
\midrule
EPEX Spot CID & L2-based baselines & 0.621 & 0.556 & 0.503 & 0.466 \\
     & L3-based models (ours) & \textbf{0.659} & \textbf{0.586} & \textbf{0.517} & \textbf{0.472} \\
\midrule
NASDAQ AAPL & L2-based baselines & 0.658 & 0.527 & 0.435 & 0.396 \\
            & L3-based models (ours) & \textbf{0.667} & \textbf{0.537} & \textbf{0.445} & \textbf{0.400} \\
\midrule
NYSE BAC & L2-based baselines & 0.733 & \textbf{0.592} & 0.438 & 0.383 \\
         & L3-based models (ours) & \textbf{0.735} & 0.589 & \textbf{0.455} & \textbf{0.393} \\
\bottomrule
\end{tabular}
\end{table}

Table~\ref{tab:predictive-summary} summarizes predictive quality across representations, and Appendix~\ref{app:full-results} reports the detailed results for every model, market and horizon. We organize the results around three findings that bear directly on price prediction under tight inference and memory budgets.
\paragraph{L3-based models improve predictive quality at fixed model capacity.}
One of our two L3-based models attains the highest mean macro-F1 in eleven of the twelve tested market-by-horizon settings. At $h=10$, our 7{,}203-parameter MBOFormer model scores a macro-F1 of 0.651 on EPEX, 0.665 on AAPL, and 0.735 on BAC data, exceeding both the 1.20M-parameter TLOB (0.617, 0.653, and 0.731) and the 3.14M-parameter MLPLOB (0.612, 0.657, and 0.732) on all three markets. Increasing TLOB from 23.9k to 1.20M parameters provides no consistent gain, while the capacity-matched MBOFormer exceeds the 7.2k-parameter TLOB in eleven of the twelve settings. Because TLOB also receives the triggering message and engineered OFI and VWAP features, this comparison suggests gains beyond providing triggering-message features and engineered order-flow summaries.

The advantage persists across forecast horizons and although the macro-F1 score declines as the horizon increases, one of our L3-based models remains best at $h=50$ and $h=100$ for every market. The temporal-context branch is not uniformly beneficial: MBOFusion improves all four AAPL results, the shorter EPEX horizons, and the longer BAC horizons, whereas MBOFormer performs better in the remaining settings. The L3 event sequence therefore accounts for the principal gain, with the context branch providing a market- and horizon-dependent refinement.
\paragraph{Predictive gains remain compatible with tight inference budgets.}
We measure forward-pass latency for all models at batch size one (one prediction per pass) on a single Apple M2 CPU thread, with models and shape-valid synthetic inputs set to FP32 precision. We report CPU rather than GPU latency because, at batch size one, the M2 GPU is slower for these small graphs (2.5\,ms for MBOFormer), as kernel-launch and host-synchronization overhead dominates the computation. After 100 warm-up calls, we pool 2{,}500 synchronized forward passes. The measurements exclude preprocessing, data transfer, model construction, and decision execution. We also report serialized state-dictionary size and FLOPs per forward pass, counted with PyTorch's \texttt{FlopCounterMode}, while estimating peak activation memory from one warmed-up forward pass by ordering the allocation and deallocation events recorded by \texttt{torch.profiler}.

\begin{table}[t]
\centering
\small
\setlength{\tabcolsep}{4pt}
\caption{Model-only batch-one inference on one Apple M2 CPU thread in float32 over 2{,}500 forwards. State is the serialized state dictionary, Act.\ is a \texttt{torch.profiler} estimate of peak live intermediate-tensor memory during one warmed-up forward pass excluding model weights and the preallocated input, and MFLOPs are per forward pass.}
\label{tab:inference}
\begin{tabular}{lrrrrrrr}
\toprule
Model & Params & State kB & Act.\ kB & MFLOPs & p50 ms & p95 ms & p99 ms \\
\midrule
MBOFormer (ours) & 7.2k & 39 & 57 & 1.091 & 0.479 & 0.552 & 0.791 \\
MBOFusion (ours) & 14.4k & 69 & 60 & 1.209 & 0.531 & 0.603 & 0.862 \\
TLOB 7.2k & 7.2k & 45 & 35 & 0.139 & 0.301 & 0.363 & 0.610 \\
TLOB 23.9k & 23.9k & 124 & 35 & 0.713 & 0.578 & 0.725 & 0.919 \\
BiNCTABL & 11.4k & 51 & 16 & 0.226 & 0.187 & 0.208 & 0.315 \\
DeepLOB & 143.9k & 612 & 512 & 70.364 & 3.468 & 4.971 & 6.800 \\
TLOB 1.20M & 1.20M & 4{,}888 & 334 & 123.646 & 2.459 & 2.810 & 3.424 \\
MLPLOB & 3.14M & 12{,}564 & 834 & 283.369 & 3.903 & 5.665 & 6.602 \\
\bottomrule
\end{tabular}
\end{table}

Table~\ref{tab:inference} distinguishes parameter efficiency from runtime efficiency. MBOFormer and MBOFusion both achieve sub-millisecond median CPU latency and their serialized state dictionaries occupy 39 and 69\,kB, respectively. We separately report activation-memory estimates in Table~\ref{tab:inference} where the figures exclude framework overhead and the state required for order-book reconstruction and feature preprocessing. By comparison, the serialized state dictionaries of the 1.20M-parameter TLOB and the 3.14M-parameter MLPLOB occupy 4.9 and 12.6\,MB. Relative to these two baselines, MBOFormer's median latency is 5.1 and 8.1 times lower, respectively, and it requires 113 and 260 times fewer FLOPs. The compact baselines present a different trade-off: the parameter-matched TLOB and BiNCTABL are faster because their shorter temporal inputs require fewer operations, but they generally attain lower predictive quality. MBOFormer therefore offers a favorable quality-efficiency trade-off rather than the lowest latency in isolation and both of our models remain below one millisecond at p99.

\paragraph{The advantage of L3-based models varies across markets.}
At $h=10$, the best L3-based result exceeds the best L2-based result by 0.038 in macro-F1 on EPEX (0.659 versus 0.621), 0.009 on AAPL (0.667 versus 0.658), and 0.002 on BAC (0.735 versus 0.733). The largest observed difference occurs on EPEX, the thinnest and slowest market in the study. However, because the datasets also differ in instrument type, trading venue, spread behavior, activity, and effective wall-clock horizon, our experiments do not identify which market characteristic explains these differences.

The equity datasets exhibit a similar descriptive ordering. One-tick bid--ask spreads occur in approximately 33\% of AAPL snapshots and 91\% of BAC snapshots, while the corresponding short-horizon L3--L2 differences are only 0.009 and 0.002 in macro-F1. This pattern is consistent with the hypothesis that event histories may be more useful in some liquidity regimes, but evidence from two instruments on different exchanges cannot safely establish spread behavior, liquidity, or information lost through snapshot aggregation as the underlying mechanism. These scores do not establish trading profitability, but they show performance well above the constant baseline and demonstrate that the same compact model family can be applied to both electricity and equity markets.

\endgroup

\section{Conclusion}
Across three markets and four horizons, our compact transformers over fine-grained L3 event history attain the highest mean macro-F1 in eleven of the twelve tested settings while achieving sub-millisecond median latency with serialized state dictionaries below 70\,kB. These results suggest that, for the markets and baselines considered here, fine-grained L3 event histories can enable compact models to match or outperform substantially larger L2-based models. Several limitations remain. The experiments cover only three markets, and the three-seed results do not establish statistical significance across test periods. Our latency measurements capture only model-forward execution on one standard laptop CPU rather than the latency-tuned production hardware used in exchange co-located trading systems. We report FLOPs rather than measured energy, and the macro-F1 scores generally do not represent trading revenue. Future work should evaluate additional instruments and market regimes, benchmark complete online pipelines on deployment hardware, and assess actual economic performance after transaction costs.

\newpage
\bibliographystyle{unsrtnat}
\bibliography{references}

\clearpage
\appendix

\section{Baseline implementation details}
\label{app:baseline-details}

Our TLOB baseline is based on the code publication \cite{berti2025tlob}. We integrated it into the common pipeline to ensure that all models use identical decision points, labels, and day splits. We added dropout, replaced layer normalization with root-mean-square normalization, adopted a more efficient attention implementation, and made the model width, depth, and input-window length configurable. The 1.20M-parameter variant processes 128 sampled L2 rows, whereas the 23.9k- and 7.2k-parameter variants process 16. We compared our adaptation with the released implementation using identical model configurations and found comparable predictive performance.

\section{Full predictive results}
\label{app:full-results}
We present our full predictive results comparison between the previous benchmarks and our two new models for the three datasets in Table~\ref{tab:full-results}. Our models attain the highest mean macro-F1 in eleven of the twelve settings, at 7{,}203 and 14{,}371 parameters against 1.20 and 3.14 million for the strongest L2-based baselines.

\begin{table}[ht]
\centering
\small
\caption{Test macro-F1, shown as mean and standard deviation over three seeds, for EPEX Spot CID, NASDAQ AAPL, and NYSE BAC. The L2-based baselines (DeepLOB, BiNCTABL, MLPLOB, and TLOB) precede our L3-based models (MBOFormer and MBOFusion) within each market.}
\label{tab:full-results}
\resizebox{\textwidth}{!}{%
\begin{tabular}{lrcccc}
\toprule
Model & Params & $h{=}10$ & $h{=}20$ & $h{=}50$ & $h{=}100$ \\
\midrule
\multicolumn{6}{l}{\textbf{EPEX Spot CID}} \\
\addlinespace[1pt]
DeepLOB           & 143.9k & $0.589_{\pm 0.018}$ & $0.508_{\pm 0.013}$ & $0.427_{\pm 0.003}$ & $0.409_{\pm 0.002}$ \\
BiNCTABL          & 11.4k & $0.596_{\pm 0.003}$ & $0.533_{\pm 0.002}$ & $0.486_{\pm 0.000}$ & $0.454_{\pm 0.001}$ \\
MLPLOB            & 3.14M & $0.612_{\pm 0.002}$ & $0.517_{\pm 0.008}$ & $0.413_{\pm 0.001}$ & $0.389_{\pm 0.007}$ \\
TLOB              & 1.20M & $0.617_{\pm 0.003}$ & $0.552_{\pm 0.003}$ & $0.503_{\pm 0.003}$ & $0.463_{\pm 0.003}$ \\
TLOB              & 23.9k & $0.621_{\pm 0.001}$ & $0.556_{\pm 0.002}$ & $0.502_{\pm 0.001}$ & $0.462_{\pm 0.002}$ \\
TLOB              & 7.2k  & $0.613_{\pm 0.003}$ & $0.550_{\pm 0.001}$ & $0.501_{\pm 0.002}$ & $0.466_{\pm 0.003}$ \\
\cmidrule(lr){1-6}
MBOFormer (ours)  & 7.2k  & $0.651_{\pm 0.001}$ & $0.577_{\pm 0.001}$ & \best{0.517}{0.001} & \best{0.472}{0.002} \\
MBOFusion (ours)  & 14.4k & \best{0.659}{0.001} & \best{0.586}{0.000} & $0.515_{\pm 0.003}$ & $0.460_{\pm 0.001}$ \\
\addlinespace[5pt]
\multicolumn{6}{l}{\textbf{NASDAQ AAPL}} \\
\addlinespace[1pt]
DeepLOB           & 143.9k & $0.641_{\pm 0.000}$ & $0.503_{\pm 0.007}$ & $0.355_{\pm 0.005}$ & $0.342_{\pm 0.003}$ \\
BiNCTABL          & 11.4k & $0.652_{\pm 0.000}$ & $0.522_{\pm 0.001}$ & $0.432_{\pm 0.001}$ & $0.388_{\pm 0.001}$ \\
MLPLOB            & 3.07M & $0.657_{\pm 0.001}$ & $0.525_{\pm 0.001}$ & $0.415_{\pm 0.005}$ & $0.344_{\pm 0.001}$ \\
TLOB              & 1.20M & $0.653_{\pm 0.001}$ & $0.525_{\pm 0.001}$ & $0.431_{\pm 0.001}$ & $0.391_{\pm 0.001}$ \\
TLOB              & 23.9k & $0.658_{\pm 0.000}$ & $0.527_{\pm 0.000}$ & $0.435_{\pm 0.000}$ & $0.396_{\pm 0.000}$ \\
TLOB              & 7.2k  & $0.656_{\pm 0.000}$ & $0.527_{\pm 0.001}$ & $0.435_{\pm 0.000}$ & $0.395_{\pm 0.001}$ \\
\cmidrule(lr){1-6}
MBOFormer (ours)  & 7.2k  & $0.665_{\pm 0.000}$ & $0.535_{\pm 0.000}$ & $0.442_{\pm 0.000}$ & $0.397_{\pm 0.000}$ \\
MBOFusion (ours)  & 14.4k & \best{0.667}{0.000} & \best{0.537}{0.000} & \best{0.445}{0.001} & \best{0.400}{0.003} \\
\addlinespace[5pt]
\multicolumn{6}{l}{\textbf{NYSE BAC}} \\
\addlinespace[1pt]
DeepLOB           & 143.9k & $0.719_{\pm 0.001}$ & $0.576_{\pm 0.003}$ & $0.404_{\pm 0.007}$ & $0.350_{\pm 0.006}$ \\
BiNCTABL          & 11.4k & $0.718_{\pm 0.000}$ & $0.579_{\pm 0.001}$ & $0.425_{\pm 0.001}$ & $0.374_{\pm 0.001}$ \\
MLPLOB            & 3.07M & $0.732_{\pm 0.001}$ & \best{0.592}{0.002} & $0.404_{\pm 0.006}$ & $0.344_{\pm 0.009}$ \\
TLOB              & 1.20M & $0.731_{\pm 0.000}$ & $0.589_{\pm 0.001}$ & $0.438_{\pm 0.005}$ & $0.383_{\pm 0.007}$ \\
TLOB              & 23.9k & $0.733_{\pm 0.001}$ & $0.591_{\pm 0.001}$ & $0.434_{\pm 0.002}$ & $0.383_{\pm 0.001}$ \\
TLOB              & 7.2k  & $0.732_{\pm 0.001}$ & $0.591_{\pm 0.002}$ & $0.434_{\pm 0.001}$ & $0.374_{\pm 0.008}$ \\
\cmidrule(lr){1-6}
MBOFormer (ours)  & 7.2k  & \best{0.735}{0.000} & $0.589_{\pm 0.001}$ & $0.439_{\pm 0.005}$ & $0.375_{\pm 0.005}$ \\
MBOFusion (ours)  & 14.4k & $0.730_{\pm 0.001}$ & $0.587_{\pm 0.001}$ & \best{0.455}{0.002} & \best{0.393}{0.002} \\
\bottomrule
\end{tabular}
}
\end{table}

\end{document}